\documentclass[twocolumn,preprintnumbers,amsmath,amssymb]{revtex4}
\usepackage{graphicx}
\usepackage{amssymb}
\usepackage{multirow}

\begin{document}

\title{Training quantum machine learning models on cloud without uploading
the data}
\author{Guang Ping He}
\email{hegp@mail.sysu.edu.cn}
\affiliation{School of Physics, Sun Yat-sen University, Guangzhou 510275, China}

\begin{abstract}
Based on the linearity of quantum unitary operations, we propose a method
that runs the parameterized quantum circuits before encoding the input data.
This enables a dataset owner to train machine learning models on quantum
cloud computation platforms, without the risk of leaking the information
about the data. It is also capable of encoding a vast amount of data
effectively at a later time using classical computations, thus saving
runtime on quantum computation devices. The trained quantum machine learning
models can be run completely on classical computers, meaning the dataset
owner does not need to have any quantum hardware, nor even quantum
simulators. Moreover, our method mitigates the encoding bottleneck by
reducing the required circuit depth from $O(2^{n})$ to $O(n)$, and relax the
tolerance on the precision of the quantum gates for the encoding. These
results demonstrate yet another advantage of quantum and quantum-inspired
machine learning models over existing classical neural networks, and broaden
the approaches to data security.
\end{abstract}

%\keywords{data security, machine learning, variational quantum circuit, federated learning, shadow model}

%(PACS needs to be changed)
%\pacs{03.67.Dd, 42.50.Ex, 03.67.Ac, 03.67.Hk, 03.65.Ud}
\maketitle

%%%%%%%%%%%%%%%%%%%%%%%%%%%%%%%%%%%%%%%%%%%%%%%%%%%%%%%%%%%%%%%%%%%%%%%%%%%%%%%%%%%%%%%%%%%%%%%%%%%%%%%%%%%%%%%%%%%%%%%%%%%%%%%%%%%%%%%%%%%%%%%%%%%%%%%%%%%%%%%%%%%%%%%%%%%%%%%%%%%%%%%%%%%%%%%%%%%%%%%%%%%%%%%%%%%%%%%%%%%%%%%%%%%%%%%%%%%%%%%%%%%%%%%%%%%%

\section{Introduction}

Data security is rated more and more important nowadays. Individuals need to
keep their privacy from misuse and abuse, companies want to protect their
intellectual property rights, and governments are concerned about the threat
to the national security. Unauthorized transfer of data may be against the
law in many places. Some countries also put strict restrictions on exporting
data abroad. On the other hand, the rapid development of artificial
intelligence technology demands a vast amount of data as input, notably in
the training of autonomous driving, intelligent healthcare systems and large
language models. Meanwhile, not every data owner has very powerful
computation devices at home, nor even in his own region. Especially, while
quantum machine learning is generally expected to be a potential powerful
tool for artificial intelligence technology, intermediate and large scale
quantum computers are only available via very few cloud platforms in certain
countries.

To solve the dilemma between the owner of data and the provider of
computational resources, here we propose a run-before-encoding method, which
can accomplish the following task. Suppose that Alice owns the dataset and
Bob holds the quantum cloud computation platform. Alice can run the quantum
circuits on Bob's platform beforehand, without encoding any data.
Based on the output received from Bob, Alice can input her data later and
calculate the cost function needed for training machine learning models
on her own local classical computer. The final trained models can also be run
on Alice's local classical computer. Since her data has never been sent to
Bob's side, it remains perfectly secure against Bob. This merit makes the
method very useful either as a standalone application or as a building block
for federated learning \cite{ml220,ml213,ml314}.

This method is backed by the linearity in quantum unitary operations, so
that it is unavailable for existing classical neural networks where the
activation functions of the neurons are nonlinear. Thus, it displays yet
another advantage of using quantum or quantum-inspired machine learning
models over existing classical counterparts.

\section{Typical features of quantum machine learning models}

Our method works for quantum circuits with the following features: (1)
measurements are performed at the last stage only, while all the rest
operations are unitary transformations, and (2) these unitary
transformations need to result in real probability amplitudes only.
Fortunately, a large portion of the variational quantum circuit (VQC)
architecture \cite{ml157,ml54,ml53,ml208,ml32} widely used for quantum
machine learning today belongs to this category, as elaborated below.

%%%%%%%%%%%%%%%%%%%%%%%%%%%%%%%%%%%%%%%%%%%%%%%%%%%%%%%%%%%%%%%%%%%%%%%%

\begin{figure*}[tbp]
\includegraphics[scale=0.85]{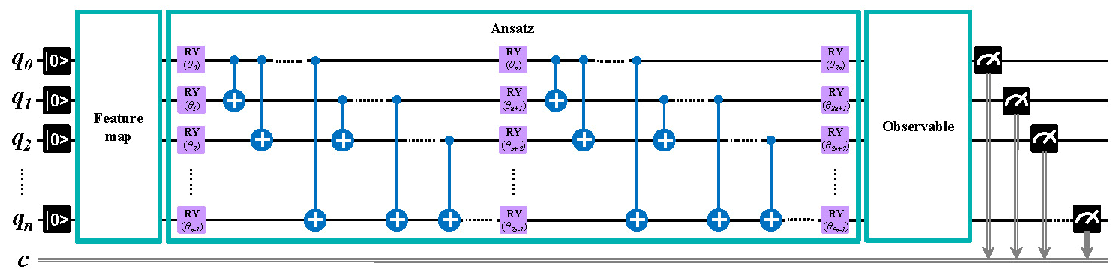}
\caption{A typical variational quantum circuit (VQC) with the \textit{RealAmplitudes} ansatz. Showed with 2-repetition and full entanglement.}
\label{fig:epsart}
\end{figure*}

%%%%%%%%%%%%%%%%%%%%%%%%%%%%%%%%%%%%%%%%%%%%%%%%%%%%%%%%%%%%%%%%%%%%%%%%

To avoid confusion, we use $\left\vert x\right\rangle _{2}$\ ($\left\vert
x\right\rangle _{10}$) to denote the quantum states where $x$ is understood
as a binary (decimal) number. Taking the 5-qubit states as an example, we
have $\left\vert 0\right\rangle _{10}=\left\vert 0\right\rangle _{2}\otimes
\left\vert 0\right\rangle _{2}\otimes \left\vert 0\right\rangle _{2}\otimes
\left\vert 0\right\rangle _{2}\otimes \left\vert 0\right\rangle
_{2}=\left\vert 00000\right\rangle _{2}$, $\left\vert 1\right\rangle
_{10}=\left\vert 00001\right\rangle _{2}$, $\left\vert 2\right\rangle
_{10}=\left\vert 00010\right\rangle _{2}$, etc. Fig. 1 shows the typical
architecture of a VQC, which consists of three modules: the feature map, the
ansatz, and the measurement of the observables. At the beginning, the state
of all the qubits on the far left are initialized as $\left\vert
0\right\rangle _{10}$. Then the feature map serves as a unitary
transformation $U_{F}(x)$, which actually consists of several 1-qubit and
2-qubit elementary unitary quantum gates, turning $\left\vert 0\right\rangle
_{10}$ into a certain quantum state $\left\vert x\right\rangle _{10}$ that
encodes the input data $x$. There are many data-encoding methods, e.g.,
dense angle encoding, general qubit encoding, amplitude encoding and general
amplitude encoding \cite{ml118,ml269}. The encoded state in all these
methods can be expressed as a linear superposition of the basis vectors.
That is, for an input data described by the $d$-dimensional feature vector $%
x=[x_{0},...,x_{d-1}]^{T}\in R^{d}$, the feature map encodes it using the
state%
\begin{equation}
\left\vert x\right\rangle _{10}=U_{F}(x)\left\vert 0\right\rangle _{10}=%
\frac{1}{C_{x}}\sum\limits_{i=0}^{d-1}f_{i}(x)\left\vert i\right\rangle
_{10},  \label{general encoding}
\end{equation}%
where $f_{i}(x)$ is a function of $x$, $\left\vert i\right\rangle _{10}$ ($%
i=0,...,d-1$, $d\leq 2^{n}$) are the basis vectors of the Hilbert space of
the $n$-qubit system, and $C_{x}$ is the normalization constant.

The ansatz has many possible structures too. A widely used one is the
\textit{RealAmplitudes} ansatz \cite{ml53,ml32}, which is the default ansatz
in Qiskit's VQC implementation. It can have many repetitions and different
entanglement constructions. The example in Fig. 1 is a 2-repetition one with
\textit{full entanglement} (see Sec. 1 of Supplementary Materials for
explanation). The effect of the entire \textit{RealAmplitudes} ansatz can be
described by a unitary transformation $U_{A}(\vec{\theta})$ with $\vec{\theta%
}$ being a set of adjustable parameters. It has the following merit. When
being applied on any $n$-qubit state%
\begin{equation}
\left\vert \psi \right\rangle _{10}=\sum\limits_{i=0}^{N-1}A_{i}\left\vert
i\right\rangle _{10},
\end{equation}%
if all $A_{i}$ ($i=0,...,N-1$, $N=2^{n}$) are real numbers, then in the
resultant state%
\begin{equation}
\left\vert \psi ^{\prime }\right\rangle _{10}=U_{A}(\vec{\theta})\left\vert
\psi \right\rangle _{10}=\sum\limits_{i=0}^{N-1}A_{i}^{\prime }\left\vert
i\right\rangle _{10},
\end{equation}%
the amplitudes $A_{i}^{\prime }$ ($i=0,...,N-1$) will remain real.

Finally, the measurement module measures a certain observable of the qubits
in a certain basis. It can always be rephrased as applying a certain unitary
transformation $U_{O}$\ on all qubits, then measuring them in the
computational basis $\{\left\vert 0\right\rangle _{10},\left\vert
1\right\rangle _{10},\left\vert 2\right\rangle _{10},...,\left\vert
N-1\right\rangle _{10}\}$. This is the only part where nonlinearity occurs
in a VQC, because for each input $x$, the final state right before the
measurement in the computational basis at the end of the VQC can be
expressed as%
\begin{equation}
\left\vert \psi _{final}(x)\right\rangle
_{10}=\sum\limits_{m=0}^{N-1}a_{m}(x)\left\vert m\right\rangle _{10},
\label{final}
\end{equation}%
where $a_{m}(x)$\ ($m=0,...,N-1$) is the normalized probability amplitude
for the basis vector $\left\vert m\right\rangle _{10}$. The probability for
finding $\left\vert m\right\rangle _{10}$ is%
\begin{equation}
p_{m}(x)=\left\vert a_{m}(x)\right\vert ^{2},  \label{pm}
\end{equation}%
which is a nonlinear function.

Once $p_{m}(x)$ is obtained, it can be used for computing the cost function
and the gradients of the adjustable parameters via classical computations
\cite{ml94,97ofML94}. This completes one epoch of training.

\section{Our method}

Finding $p_{m}(x)$ is the key of the training process. In previous research,
it is done simply by running the VQC using $x$ as input. If Alice wants to
use Bob's quantum cloud computation platform, she has to send $x$ to Bob.
But now we are interested in how to obtain $p_{m}(x)$ without sending $x$
out of Alice's site. To this end, notice that the relationship between the
initial state $\left\vert 0\right\rangle _{10}$, the state $\left\vert
x\right\rangle _{10}$ after applying the feature map, and the final state $%
\left\vert \psi _{final}(x)\right\rangle _{10}$ before the measurement in
the computational basis is%
\begin{equation}
\left\vert \psi _{final}(x)\right\rangle _{10}=U_{O}U_{A}(\vec{\theta}%
)U_{F}(x)\left\vert 0\right\rangle _{10}=U_{O}U_{A}(\vec{\theta})\left\vert
x\right\rangle _{10}.
\end{equation}%
Combining with Eqs. (\ref{general encoding}) and (\ref{final}), we yield%
\begin{equation}
\sum\limits_{m=0}^{N-1}a_{m}(x)\left\vert m\right\rangle
_{10}=\sum\limits_{i=0}^{d-1}\frac{f_{i}(x)}{C_{x}}U_{O}U_{A}(\vec{\theta}%
)\left\vert i\right\rangle _{10}.
\end{equation}%
Expanding $U_{O}U_{A}(\vec{\theta})\left\vert i\right\rangle _{10}$\ in the
computational basis as%
\begin{equation}
U_{O}U_{A}(\vec{\theta})\left\vert i\right\rangle
_{10}=\sum\limits_{m=0}^{N-1}b_{m}(i)\left\vert m\right\rangle _{10},
\label{basis}
\end{equation}%
thus we have%
\begin{equation}
a_{m}(x)=\sum\limits_{i=0}^{d-1}\frac{f_{i}(x)}{C_{x}}b_{m}(i)
\label{deduced_amplitude}
\end{equation}%
for $m=0,...,N-1$. It can be expressed in a more compact form using matrix
product as%
\begin{equation}
\vec{a}(x)=B\vec{f}(x),  \label{matrix product}
\end{equation}%
where $\vec{a}(x)$ is a $N$-dimensional vector defined as%
\begin{equation}
\vec{a}(x)=[a_{0}(x),...,a_{m}(x),...,a_{N-1}(x)]^{T},
\end{equation}%
$\vec{f}(x)$ is a $d$-dimensional vector defined as%
\begin{equation}
\vec{f}(x)=\frac{1}{C_{x}}[f_{0}(x),...,f_{i}(x),...,f_{d-1}(x)]^{T},
\end{equation}%
and $B$ is a $N\times d$ matrix defined as%
\begin{equation}
B=\left[
\begin{array}{ccccc}
b_{0}(0) & \cdots & b_{0}(i) & \cdots & b_{0}(d-1) \\
\vdots & \ddots & \vdots &  & \vdots \\
b_{m}(0) & \cdots & b_{m}(i) & \cdots & b_{m}(d-1) \\
\vdots &  & \vdots & \ddots & \vdots \\
b_{N-1}(0) & \cdots & b_{N-1}(i) & \cdots & b_{N-1}(d-1)%
\end{array}%
\right] ,  \label{amp matrix}
\end{equation}%
which is actually the first $d$ columns of the $N\times N$\ matrix
representation of $U_{O}U_{A}(\vec{\theta})$.

Eqs. (\ref{basis}) and (\ref{matrix product}) give birth to the central idea
of our method. The former implies that $b_{m}(i)$\ ($m=0,...,N-1$) are the
probability amplitudes of the state right before the final measurement of
the VQC when the basis vector $\left\vert i\right\rangle _{10}$ ($%
i=0,...,d-1 $) is input directly to the ansatz, skipping the feature map.
Notably, $b_{m}(i)$ is not a function of $x$. Consequently, it can be
calculated without transferring $x$ from Alice to Bob. The latter equation
means that Alice can try to obtain all $b_{m}(i)$\ ($m=0,...,N-1$, $%
i=0,...,d-1$) (i.e., the matrix $B$) using Bob's quantum computation
platform first. Then she can decide on the encoding method of her feature
map at a later time, which determines the form of the function $f_{i}(x)$
and the normalization constant $C_{x}$. Finally, she substitutes the values
of $b_{m}(i)$ received from Bob into Eq. (\ref{matrix product}) to compute $%
\vec{a}(x)$ and obtain all $p_{m}(x)=\left\vert a_{m}(x)\right\vert ^{2}$ ($%
m=0,...,N-1$), which is simple arithmetic that can be done on her own
classical computers without even using quantum simulators.

However, the probability amplitude $b_{m}(i)$ is not an observable. Although
some quantum simulators can compute it directly, on real quantum hardwares
this cannot be done. Instead, real quantum computers can output the
probability $p_{m}(\left\vert i\right\rangle _{10})=\left\vert
b_{m}(i)\right\vert ^{2}$ only. Since $b_{m}(i)$ can be a complex number in
general, it can be expressed as%
\begin{equation}
b_{m}(i)=s_{m}(i)\sqrt{p_{m}(\left\vert i\right\rangle _{10})},
\label{amp vs prob}
\end{equation}%
where $s_{m}(i)$\ could be any complex number of unit modulus. Fortunately,
as mentioned above, the VQC studied here has the feature that the
probability amplitudes $b_{m}(i)$ of the final state before the last
measurement are always real. In this case, the possible choices\ are reduced
to either $s_{m}(i)=1$ or $s_{m}(i)=-1$. Still, $p_{m}(\left\vert
i\right\rangle _{10})$ alone is insufficient for determining $s_{m}(i)$\
uniquely.

%%%%%%%%%%%%%%%%%%%%%%%%%%%%%%%%%%%%%%%%%%%%%%%%%%%%%%%%%%%%%%%%%%%%%%%%

\begin{figure*}[tbp]
\includegraphics[scale=0.75]{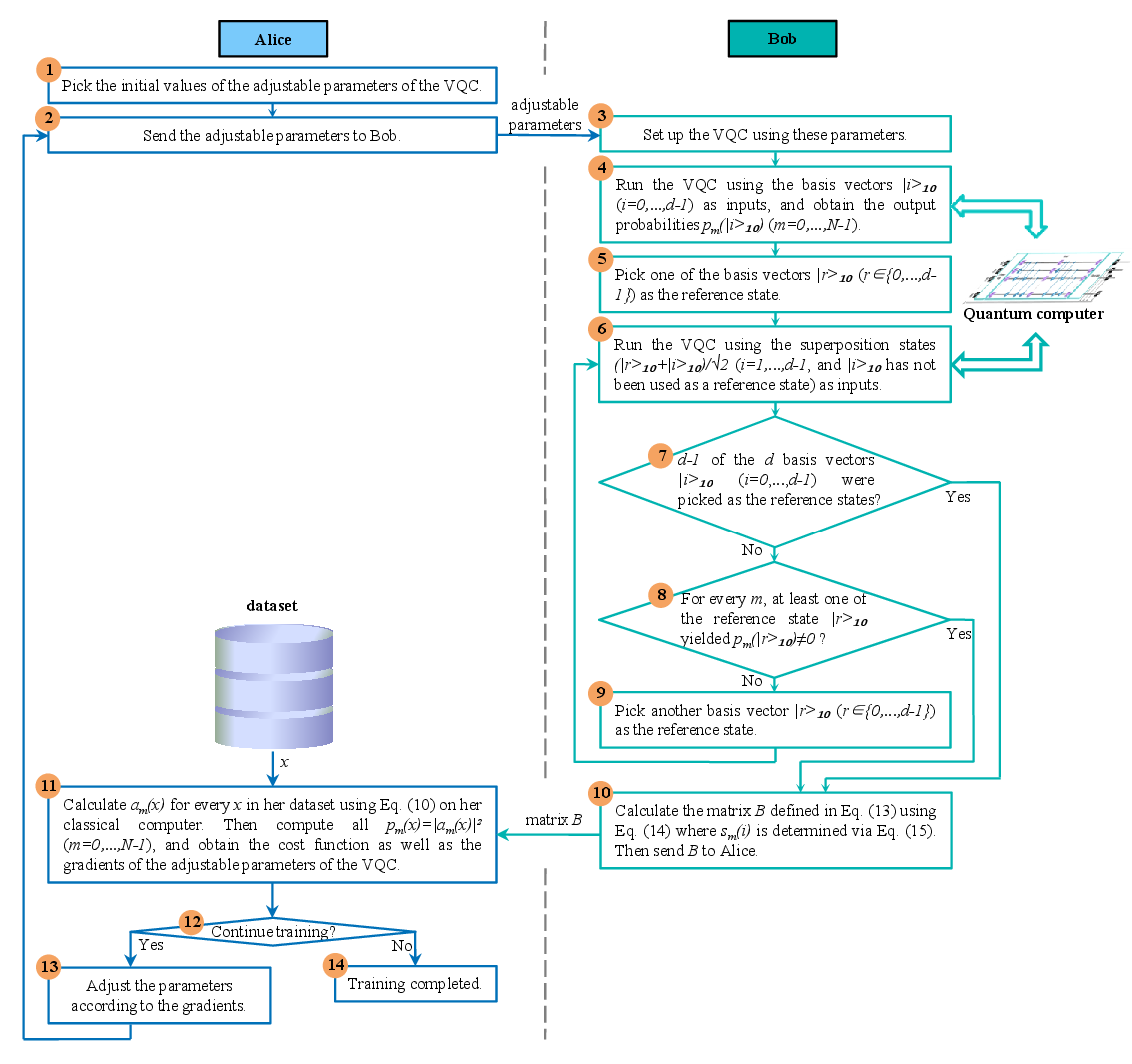}
\caption{Flow chart of our run-before-encoding method.}
\label{fig:epsart}
\end{figure*}

%%%%%%%%%%%%%%%%%%%%%%%%%%%%%%%%%%%%%%%%%%%%%%%%%%%%%%%%%%%%%%%%%%%%%%%%

To circumvent this difficulty, we have the following trick. First, by using
the basis vectors $\left\vert i\right\rangle _{10}$ ($i=0,...,d-1$. Note
that Eq. (\ref{deduced_amplitude}) implies that $i=d,...,N-1$\ are not
needed) as inputs to the ansatz (skipping the feature map), Alice asks Bob
to compute $p_{m}(\left\vert i\right\rangle _{10})$ ($m=0,...,N-1$, $%
i=0,...,d-1$) on his quantum platform. Next, Bob picks one of the basis
vectors as the reference state $\left\vert r\right\rangle _{10}$ ($r\in
\{0,...,d-1\}$), such that most of the output $p_{m}(\left\vert
r\right\rangle _{10})$ ($m=0,...,N-1$) are nonzero. With the states $%
(\left\vert r\right\rangle _{10}+\left\vert i\right\rangle _{10})/\sqrt{2}$\
($i=0,...,d-1$, $i\neq r$) as inputs to the ansatz, he computes $%
p_{m}((\left\vert r\right\rangle _{10}+\left\vert i\right\rangle _{10})/%
\sqrt{2})$ ($m=0,...,N-1$). Then as proven in Sec. 2 of Supplementary
Materials, the relative relationship between the sign of $s_{m}(i)$ and $%
s_{m}(r)$ can be determined as%
\begin{equation}
s_{m}(i)=\left\{
\begin{array}{cc}
+s_{m}(r), & \left(
\begin{array}{c}
2p_{m}\left( \frac{\left\vert r\right\rangle _{10}+\left\vert i\right\rangle
_{10}}{\sqrt{2}}\right)  \\
>p_{m}(\left\vert r\right\rangle _{10})+p_{m}(\left\vert i\right\rangle
_{10})%
\end{array}%
\right) . \\
-s_{m}(r), & \left(
\begin{array}{c}
2p_{m}\left( \frac{\left\vert r\right\rangle _{10}+\left\vert i\right\rangle
_{10}}{\sqrt{2}}\right)  \\
<p_{m}(\left\vert r\right\rangle _{10})+p_{m}(\left\vert i\right\rangle
_{10})%
\end{array}%
\right) .%
\end{array}%
\right.   \label{relative sign}
\end{equation}%
Consequently, we can write%
\begin{equation}
s_{m}(i)=\sigma _{m}(i)s_{m}(r)
\end{equation}%
where $\sigma _{m}(i)=s_{m}(i)/s_{m}(r)$\ is known. Combining with Eqs. (\ref%
{deduced_amplitude}) and (\ref{amp vs prob}), we find%
\begin{equation}
a_{m}(x)=s_{m}(r)\sum\limits_{i=0}^{d-1}\frac{f_{i}(x)}{C_{x}}\sigma _{m}(i)%
\sqrt{p_{m}(\left\vert i\right\rangle _{10})}.
\end{equation}%
Although $s_{m}(r)$\ remains unknown, the fact that $(s_{m}(r))^{2}=1$
ensures that we are able to calculate the probability%
\begin{equation}
p_{m}(x)=\left\vert a_{m}(x)\right\vert ^{2}=\left( \sum\limits_{i=0}^{d-1}%
\frac{f_{i}(x)}{C_{x}}\sigma _{m}(i)\sqrt{p_{m}(\left\vert i\right\rangle
_{10})}\right) ^{2}.
\end{equation}%
To sum up, by using the state vectors $\left\vert i\right\rangle _{10}$ ($%
i=0,...,d-1$) and $(\left\vert r\right\rangle _{10}+\left\vert
i\right\rangle _{10})/\sqrt{2}$\ ($i\neq r$) as inputs (i.e., $d+(d-1)$
states in total), which contain no information on the dataset $x$, Alice can
compute the probabilities $p_{m}(\left\vert i\right\rangle _{10})$ and $%
p_{m}((\left\vert r\right\rangle _{10}+\left\vert i\right\rangle _{10})/%
\sqrt{2})$ ($m=0,...,N-1$) on Bob's quantum platform, and calculate $p_{m}(x)
$ on her own classical computer later to complete the training of her
machine learning model.

Note that if the probability $p_{m}(\left\vert r\right\rangle _{10})$ for a
certain $m$\ turns out to be zero when $\left\vert r\right\rangle _{10}$ is
chosen as the reference state, then Eq. (\ref{relative sign}) will be unable
to determine the relative sign of $s_{m}(i)$ ($i\neq r$) because $%
2p_{m}((\left\vert r\right\rangle _{10}+\left\vert i\right\rangle _{10})/%
\sqrt{2})=p_{m}(\left\vert r\right\rangle _{10})+p_{m}(\left\vert
i\right\rangle _{10})$. In this case, Bob needs to pick a second reference
state $\left\vert r^{\prime }\right\rangle _{10}$ ($r^{\prime }\neq r$) such
that $p_{m}(\left\vert r^{\prime }\right\rangle _{10})\neq 0$ for this
specific $m$, and repeat the above procedure by using $(\left\vert r^{\prime
}\right\rangle _{10}+\left\vert i\right\rangle _{10})/\sqrt{2}$\ ($%
i=0,...,d-1$, $i\neq r^{\prime }$ and $i\neq r$) as inputs, and run the VQC
for an additional $d-2$\ times to determine $s_{m}(i)$. If there are still
some other $m$ for which both $p_{m}(\left\vert r\right\rangle _{10})$ and $%
p_{m}(\left\vert r^{\prime }\right\rangle _{10})$ are zero, then he should
pick a third reference state $\left\vert r^{\prime \prime }\right\rangle
_{10}$ ($r^{\prime \prime }\in \{0,...,d-1\}$, $r^{\prime \prime }\neq r$, $%
r^{\prime \prime }\neq r^{\prime }$) and run the VQC using $(\left\vert
r^{\prime \prime }\right\rangle _{10}+\left\vert i\right\rangle _{10})/\sqrt{%
2}$\ ($i=1,...,d-1$, $i\neq r$, $i\neq r^{\prime }$, $i\neq r^{\prime \prime
}$) as inputs ... , until: (a) For each $m$, at least one of $%
p_{m}(\left\vert r\right\rangle _{10})$, $p_{m}(\left\vert r^{\prime
}\right\rangle _{10})$, $p_{m}(\left\vert r^{\prime \prime }\right\rangle
_{10})$, ... is nonzero; or (b) $d-1$ of all the $d$ basis vectors $%
\left\vert i\right\rangle _{10}$ ($i=0,...,d-1$) were picked as the
reference states. That is, the VQC needs to be run for $%
d+(d-1)+(d-2)+...+1=(d+1)d/2$ times at the most. Note that in case (b), it
does not matter even if there exists some certain $m$ for which the
probabilities $p_{m}$ of all reference states are zero. This is because the
value of $a_{m}(x)$ for any $x$ will also be zero for the same $m$, as
indicated by Eq. (\ref{deduced_amplitude}). Luckily, in our experiments to
be reported below, many sets of the adjustable parameters for the VQC with
randomly chosen values were tried and we always picked $\left\vert
r\right\rangle _{10}=\left\vert 0\right\rangle _{10}$\ as the only reference
state, and it turned out that all the probabilities $p_{m}(\left\vert
r\right\rangle _{10})$ ($m=0,...,N-1$) for this $\left\vert r\right\rangle
_{10}$ are nonzero. Therefore, running the VQC for $d+(d-1)$ times in total
is already sufficient in our experiments.

For clarity, the flow chart of the above procedure is illustrated in Fig. 2.

\section{Advantages}

An obvious and foremost advantage of our method is that the security of
Alice's dataset $x$ is well-protected. As can be seen from steps (3)--(10)
of the flow chart, Bob runs the VQC based on the parameters received from
Alice using only the basis vectors and their superpositions as inputs. None
of the data $x$ was encoded in these states, nor being sent to Bob. In fact,
the relationship between $x$ and Bob's output may even remain unknown to
Alice herself at this stage, because she does not need to decide on the
encoding method and the form of the cost function until step (11).
Therefore, it is clear that Bob stands no chance to learn $x$ and the cost
function.

Secondly, our method may save the runtime on quantum devices when the
amount of data is huge. For instance, consider the best case where all the
output $p_{m}(\left\vert r\right\rangle _{10})$ ($m=0,...,N-1$)
corresponding to the first reference state $\left\vert r\right\rangle _{10}$
are nonzero. To obtain the matrix $B$, the VQC needs to be run for $%
d+(d-1)=2d-1$ input states, regardless the size of the dataset. Once $B$ is
obtained, the probability distributions corresponding to all $x$ in the
dataset can be calculated via Eq. (\ref{matrix product}) on classical
computers. In contrast, when using the old method where the VQC needs to be
run for each input $x$ one by one, evaluating a dataset with $s$ data points
has to call for the VQC $s$ times. Therefore, our method takes a much less
occupancy time on quantum devices than the old one does for any dataset
satisfying $s>>2d-1$. As an example, the MNIST dataset \cite{MNIST,MNIST2}
widely used in machine learning research has a dimension $d=784$, whereas
the number of training data is $s=50000$. In this case, our method will be
significantly faster. On quantum cloud platforms charging by the runtime,
less occupation of the quantum devices also means less cost.

Moreover, our method can help mitigating the encoding bottleneck \cite{ml230}%
. In many encoding methods including the \textit{amplitude encoding} \cite%
{ml118}, encoding a data vector $x$ using $n$ qubits generally requires a
circuit depth of the order $O(2^{n})$ \cite{ml81,ml273}, except for certain
sparse vectors. But in our method, the inputs to the VQC are merely basis
vectors and simple superpositions of two basis vectors, which are indeed sparse. As proven in Sec. 3
of Supplementary Materials, in average, encoding a single basis vector takes
only $n/2$ single-qubit NOT gates, while encoding the superpositions of two
basis vectors takes $n/2$ single-qubit NOT gates and $n/2$ two-qubit
controlled-NOT gates.

It could also relax the tolerance on the precision of the quantum gates for
the encoding. For example, suppose that for an input data vector $x$, there
is $x_{i}=0.060465$, while for the same $i$, there is $x_{i}^{\prime
}=0.060466$ for another input $x^{\prime }$. When encoding using the old
method, the quantum gates may need to be adjusted very precisely to show the
difference. But in our method, the inputs are merely the basis vectors and
their simple superpositions. Then with the same matrix $B$ obtained from
these inputs, the output of data $x$ and $x^{\prime }$ are computed from Eq. (\ref{matrix product}) using classical arithmetic. Thus, their difference will
remain visible even if the quantum gates cannot be adjusted to manifest very
subtle variation.

In addition, after the training of the quantum model is completed, Alice
will no longer need a quantum computer to run it. She can simply use the
matrix $B$ obtained in the final epoch of the training as the machine
learning model, and applies it on any new input data $x$ by computing Eq. (%
\ref{matrix product}) on her local classical computer. That is, it serves
like a classical shadow model \cite{ml355,ml349} of the VQC. But the
difference is that existing shadow models\ usually are merely approximations
which are $\delta $-close to the original quantum learning models, while
ours works exactly as the quantum one.

\section{Experimental results}

We tested the runtime of our method experimentally using a $10$-qubit VQC,
with the training data in the MNIST dataset as inputs. Details of the VQC
are provided in Sec. 1 of Supplementary Materials.

%%%%%%%%%%%%%%%%%%%%%%%%%%%%%%%%%%%%%%%%%%%%%%%%%%%%%%%%%%%%%%%%%%%%%%%%

\begin{figure}[tbp]
\includegraphics[scale=0.78]{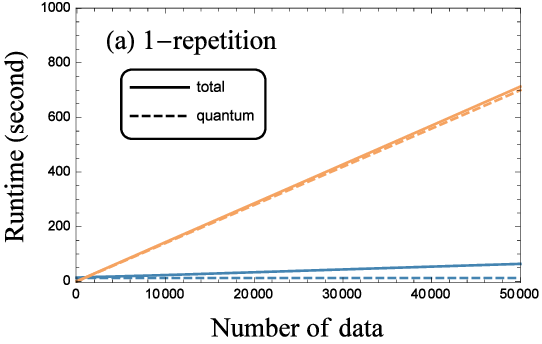} %\bigskip
\includegraphics[scale=0.78]{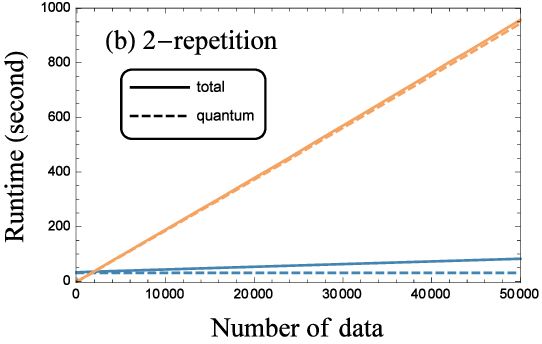}
\caption{Runtime of the VQC as a function of the number of input data, where
the ansatz of the VQC contains (a) 1 repetition, and (b) 2 repetitions. The
blues and orange lines are the results of our method and the old method,
respectively. The solid lines are the time spent in total, while the dashed
lines are the time spent on quantum devices.}
\label{fig:epsart}
\end{figure}

%%%%%%%%%%%%%%%%%%%%%%%%%%%%%%%%%%%%%%%%%%%%%%%%%%%%%%%%%%%%%%%%%%%%%%%%

Fig. 3(a) shows the comparison between the average runtime of our model and
that of the old method when the \textit{RealAmplitudes} ansatz contains $1$
repetition. (The results of the cost function turn out to be exactly the
same, just as we could expect from the above equations. Therefore, it is not
shown in the figure. Detailed values can be found in Data Availability
section.) It is found that our method is slower when the number of data is
small, but significantly faster than the old method when the number of data
exceeds $d+(d-1)=1567$. This is because our method needs to run the VQC
using the basis vectors $\left\vert i\right\rangle _{10}$ ($i=0,...,d-1$)
and the superpositions $(\left\vert 0\right\rangle _{10}+\left\vert
i\right\rangle _{10})/\sqrt{2}$\ ($i=1,...,d-1$) as inputs first, before
computing the cost function for the first data. After this is done, the VQC
is no longer needed, so that the time spent on quantum devices remains at $%
12.343$ second unchanged, regardless the number of data, as shown by the
blue dashed line. The post-processing time spent on the classical computer
for calculating the cost function grows linearly as the number of data
increases, resulting in a total runtime ranging from $13.682$ seconds (for $%
1 $ data) to $64.019$ seconds (for $50000$ data).

On the contrary, in the old method, the VQC needs to be run for every data
one by one. Thus, the total runtime is only $0.102$ seconds for finding the
output for the first data, including $0.068$ seconds spent on the quantum
circuit. But then it grows linearly and is significantly slower than
our method as the number of data increases. To achieve the cost function for
$50000$ data, the total runtime is $713.064$ seconds, including $699.879$
seconds spent on running the VQC.

In Fig. 3(b), the same comparison is repeated with the ansatz of VQC
increased to 2 repetitions. The results show that the time spent on quantum devices running the VQC increases in both methods, while the
post-processing time on the classical computer remains basically the same.
Consequently, as our method needs to run the VQC for $1567$ input states
only, the total runtime ranges from $33.091$ seconds (for $1$ data) to $%
82.701$ seconds (for $50000$ data), with the time spent on quantum devices
stays at $30.805$ seconds regardless the number of data. On the contrary,
the total runtime of the old method ranges from $0.118$ seconds (for $1$
data) to $957.232$ seconds (for $50000$ data), including the time spent on quantum devices ranging from $0.085$ seconds (for $1$ data) to $944.102$
seconds (for $50000$ data).

To sum up, these experiments verify that our method is faster than the old
method when the number of data is large, and the speed-up becomes more
significant with the increase of the depth of the quantum circuit. It also
takes much less occupancy time on quantum devices.

\section{Further improvements for better security}

In our method, though Bob does not know the value of $x$, he knows the
dimension $d$ of $x$, because he is required to run the VQC for $d$ of the
basis vectors and their superpositions. He also knows the adjustable
parameters of the VQC because Alice sent them to him. If these problems are
the major concern while increasing the runtime on quantum devices is
acceptable, we can have the following fix.

To hide the dimension of the data $x$, let us take the MNIST dataset as an
example. Originally, Alice needs only to run the VQC for $d=784$ basis
vectors instead of all the $N=2^{10}=1024$ basis vectors of $n=10$ qubits.
But if Alice wants to prevent Bob from knowing that $d=784$, she can ask Bob
to run the VQC for all the $1024$ basis vectors instead. Moreover, she can
even ask Bob to run a VQC with more than $10$ qubits, which makes it even
harder for Bob to guess the actual dimension of her data.

If Alice does not want Bob to know her choice of the adjustable parameters,
she can also generate more sets of these parameters whose values are chosen
differently by purpose, and ask Bob to run the VQC with these sets. This is
like hiding a tree in a forest, so that Bob cannot be sure which set of the
parameters is the one that Alice actually uses in her machine learning model.

\section{More discussion on the advantage}

In conclusion, we propose a run-before-encoding method so that a dataset
owner Alice can train machine learning models on Bob's quantum
% cloud computation
platform, with the primary advantage that the dataset is kept
perfectly secret from Bob. Note that it was pointed out in Ref. \cite{ml69}
that \textquotedblleft almost all branches of quantum machine-learning
research have been heavily framed by the question of `beating' classical
machine learning in some figure of merit\textquotedblright , but
\textquotedblleft a number of `positive' results have been put forward that
either `prove' theoretically or `show' empirically that quantum computers
are better at something\textquotedblright .
%According to Ref. \cite{ml353},
%it was even stated at Google I/O 2024 that \textquotedblleft no one has yet
%demonstrated a clear quantum advantage for machine learning on classical
%data\textquotedblright .
But our method works for any quantum circuit where all the operations in the
middle are unitary while the measurements are performed at the very end, and
the unitary operations always result in real probability amplitudes. These
features are not available in existing classical machine learning models but
they are not uncommon for quantum ones. Thus, the advantage of our method is
neither empirical nor biased towards certain tasks.

Nevertheless, this advantage can not only be achieved on real quantum
computers, but also persist if Bob runs quantum simulators on classical
computers. That is, though it is an advantage over all \textit{existing}
classical neural networks, strictly speaking, it may not seem appropriate to
be regarded as an exclusive quantum advantage. We feel that it should be
better considered as a \textquotedblleft quantum-inspired\textquotedblright\
advantage. This result also suggests that quantum-inspired classical neural
networks, i.e., redesigned classical neural networks by replacing the neuron
associated by existing choices of nonlinear classical activation functions
(e.g., the sigmoid, ReLU and Tanh functions) with unitary linear
transformations while nonlinearity occurs only at the very end, should
deserve more research interests since it could make full use of the
advantage in the near term.

\bigskip \textbf{Acknowledgments}

This work was supported in part by Guangdong Basic and Applied Basic
Research Foundation under Grant No. 2019A1515011048. The experiments were
supported in part by National Supercomputer Center in Guangzhou, with the
help from Qing Liu and the technical support team at National Supercomputer
Center. The author also thanks Zi-Yuan Dong for valuable discussions.

%\bigskip \textbf{Author contributions}
%
%G.P.H. researched and wrote this paper solely.
%
%\bigskip \textbf{Competing interests}
%
%The author declares no competing interests.

\bigskip \textbf{Data availability}

The raw experimental data of this study are publicly available at
https://github.com/gphehub/rbe2308.

\bigskip \textbf{Code availability}

The software codes for generating the experimental data are publicly
available at https://github.com/gphehub/rbe2308.

\newpage \setcounter{equation}{0}

%\appendix

\section*{Supplementary Materials}

\subsection*{1. Experiment details}

In our experiments, the feature map adopted the \textit{amplitude encoding}
method \cite{ml118}, i.e., a data vector $x=[x_{0},...,x_{d-1}]^{T}$ is
encoded as the state%
\begin{equation}
\left\vert x\right\rangle _{10}=\frac{1}{C_{x}}\sum\limits_{i=0}^{d-1}x_{i}%
\left\vert i\right\rangle _{10}.
\end{equation}

The \textit{RealAmplitudes} ansatz with \textit{full entanglement} was used
in our VQC. Both 1-repetition and 2-repetition were tested. Fig. 1 shows the
2-repetition architecture. It starts from a RY gate on each qubits, which
introduces an independent rotation angle $\theta $\ about the $y$-axis,
turning a qubit state $\cos \gamma \left\vert 0\right\rangle _{2}+\sin
\gamma \left\vert 1\right\rangle _{2}$\ into $\cos \left( \gamma +\frac{%
\theta }{2}\right) \left\vert 0\right\rangle _{2}+\sin \left( \gamma +\frac{%
\theta }{2}\right) \left\vert 1\right\rangle _{2}$. Then several CX
(controlled-NOT) gates are placed between every pairs of the qubits to make them entangled,
followed by another sets of RY gates on every qubit. This completes one
repetition of the \textit{RealAmplitudes} ansatz. Adding a second repetition
means adding another sets of CX gates and RY gates. The values of $\vec{%
\theta}=(\theta _{0},\theta _{1},\theta _{2},...)$\ of the RY gates are the
adjustable parameters of the ansatz.

To avoid the result being biased by the status of the VQC, the adjustable
parameters were chosen randomly and uniformly distributed over the interval $%
[0,\pi )$. With the same set of the adjustable parameters, we ran the VQC
using Qiskit's quantum simulator with both our method and the old method
(where each data point in the MNIST dataset is input into the VQC directly).
The experiment was repeated for $10$ runs, each of which used a different
set of the adjustable parameters.

The form of the cost function used for each input data $x$ is the cross
entropy \cite{ml94}%
\begin{equation}
L_{CE}(\lambda )=-\frac{1}{n}\sum_{k=1}^{n}\lambda _{k}\log (g_{k}),
\end{equation}%
where $\lambda =(\lambda _{1},...,\lambda _{n})$ denotes the label of the
input $x$ in the form of one-hot encoding \cite{97ofML94}, $g_{k}$ ($%
k=1,...,n$) denotes the probability for the $k$th qubit to be found as $%
\left\vert 1\right\rangle _{2}$ in the measurement, with $n=10$ being the
number of qubits. According to Eqs. (\ref{final}) and (\ref{pm}) of the main text, $g_{k}$
can be obtained by summing over any $p_{m}(x)$ where the state of the $k$th
qubit in $\left\vert m\right\rangle _{10}$ is $\left\vert 1\right\rangle
_{2} $.

We recorded the runtime of one epoch of training as a function of the number
of data being evaluated. It was found that the runtime varies only slightly
in different runs, indicating that it is less affected by the values of the
adjustable parameters. Therefore, we took the average of the runtime of these runs as the final result to draw Fig.3. The standard deviations were supplied in the link provided
in Data Availability section, which also includes the values of all
adjustable parameters and the runtime in each run.

All experiments were run on a personal laptop with a $2.8$ GHz quad-core
Intel i7-3840QM processor and $16$ GB $1,600$ MHz DDR3 memory.

\subsection*{2. Determining the sign of the probability amplitudes}

In our method, at first when the basis vectors $\left\vert i\right\rangle
_{10}$ ($i=0,...,d-1$) are used as inputs to the ansatz, from Eqs. (\ref%
{basis}) and (\ref{amp vs prob}) of the main text we yield%
\begin{equation}
U_{O}U_{A}(\vec{\theta})\left\vert i\right\rangle
_{10}=\sum\limits_{m=0}^{N-1}s_{m}(i)\sqrt{p_{m}(\left\vert i\right\rangle
_{10})}\left\vert m\right\rangle _{10}.  \label{stage1}
\end{equation}%
Secondly, when the states $(\left\vert r\right\rangle _{10}+\left\vert
i\right\rangle _{10})/\sqrt{2}$\ ($i=0,...,d-1$, $i\neq r$) are used as
inputs, similar to the above equation, now we have%
\begin{eqnarray}
&&U_{O}U_{A}(\vec{\theta})\left( \frac{\left\vert r\right\rangle
_{10}+\left\vert i\right\rangle _{10}}{\sqrt{2}}\right)
=\sum\limits_{m=0}^{N-1}\left( s_{m}\left( \frac{\left\vert r\right\rangle
_{10}+\left\vert i\right\rangle _{10}}{\sqrt{2}}\right) \right.   \nonumber
\\
&&\left. \cdot \sqrt{p_{m}\left( \frac{\left\vert r\right\rangle
_{10}+\left\vert i\right\rangle _{10}}{\sqrt{2}}\right) }\left\vert
m\right\rangle _{10}\right) .  \label{stage2}
\end{eqnarray}%
Due to the linearity of the unitary transformation $U_{O}U_{A}(\vec{\theta})$%
, the left-hand side of this equation can be expressed using the above Eq. (\ref%
{stage1}) as%
\begin{eqnarray}
&&U_{O}U_{A}(\vec{\theta})\left( \frac{\left\vert r\right\rangle
_{10}+\left\vert i\right\rangle _{10}}{\sqrt{2}}\right)   \nonumber \\
&=&\frac{1}{\sqrt{2}}(U_{O}U_{A}(\vec{\theta})\left\vert r\right\rangle
_{10}+U_{O}U_{A}(\vec{\theta})\left\vert i\right\rangle _{10})  \nonumber \\
&=&\sum\limits_{m=0}^{N-1}\frac{1}{\sqrt{2}}(s_{m}(r)\sqrt{p_{m}(\left\vert
r\right\rangle _{10})}  \nonumber \\
&&+s_{m}(i)\sqrt{p_{m}(\left\vert i\right\rangle _{10})})\left\vert
m\right\rangle _{10}.
\end{eqnarray}%
Comparing with the right-hand side of Eq. (\ref{stage2}), we find%
\begin{eqnarray}
&&\frac{1}{\sqrt{2}}\left( s_{m}(r)\sqrt{p_{m}(\left\vert r\right\rangle
_{10})}+s_{m}(i)\sqrt{p_{m}(\left\vert i\right\rangle _{10})}\right)
\nonumber \\
&=&s_{m}\left( \frac{\left\vert r\right\rangle _{10}+\left\vert
i\right\rangle _{10}}{\sqrt{2}}\right) \sqrt{p_{m}\left( \frac{\left\vert
r\right\rangle _{10}+\left\vert i\right\rangle _{10}}{\sqrt{2}}\right) }.
\nonumber \\
&&
\end{eqnarray}%
Recall that $s_{m}(i)=\pm 1$ which gives $(s_{m}(i))^{2}=1$ and $%
1/s_{m}(i)=s_{m}(i)$\ (so do $s_{m}(r)$\ and $s_{m}((\left\vert
r\right\rangle _{10}+\left\vert i\right\rangle _{10})/\sqrt{2})$), there is
\begin{eqnarray}
&&\left( \frac{1}{\sqrt{2}}\left( s_{m}(r)\sqrt{p_{m}(\left\vert
r\right\rangle _{10})}+s_{m}(i)\sqrt{p_{m}(\left\vert i\right\rangle _{10})}%
\right) \right) ^{2}  \nonumber \\
&=&p_{m}\left( \frac{\left\vert r\right\rangle _{10}+\left\vert
i\right\rangle _{10}}{\sqrt{2}}\right) .
\end{eqnarray}%
Therefore%
\begin{eqnarray}
s_{m}(i) &=&\frac{1}{2\sqrt{p_{m}(\left\vert r\right\rangle
_{10})p_{m}(\left\vert i\right\rangle _{10})}}\left( 2p_{m}\left( \frac{%
\left\vert r\right\rangle _{10}+\left\vert i\right\rangle _{10}}{\sqrt{2}}%
\right) \right.   \nonumber \\
&&-(p_{m}(\left\vert r\right\rangle _{10})+p_{m}(\left\vert i\right\rangle
_{10})))\cdot s_{m}(r).
\end{eqnarray}%
Since $\sqrt{p_{m}(\left\vert r\right\rangle _{10})p_{m}(\left\vert
i\right\rangle _{10})}>0$, we arrive at%
\begin{equation}
s_{m}(i)=\left\{
\begin{array}{cc}
+s_{m}(r), & \left(
\begin{array}{c}
2p_{m}\left( \frac{\left\vert r\right\rangle _{10}+\left\vert i\right\rangle
_{10}}{\sqrt{2}}\right)  \\
>p_{m}(\left\vert r\right\rangle _{10})+p_{m}(\left\vert i\right\rangle
_{10})%
\end{array}%
\right) . \\
-s_{m}(r), & \left(
\begin{array}{c}
2p_{m}\left( \frac{\left\vert r\right\rangle _{10}+\left\vert i\right\rangle
_{10}}{\sqrt{2}}\right)  \\
<p_{m}(\left\vert r\right\rangle _{10})+p_{m}(\left\vert i\right\rangle
_{10})%
\end{array}%
\right) .%
\end{array}%
\right.
\end{equation}%
Thus, Eq. (\ref{relative sign}) in Sec. III is proven.

\subsection*{3. Number of quantum gates needed for encoding the input
vectors of our method}

Here we prove that encoding the basis vectors and the simple superposition
states $(\left\vert r\right\rangle _{10}+\left\vert i\right\rangle _{10})/%
\sqrt{2}$\ takes $O(n)$\ quantum gates in average only, thus mitigating the
encoding bottleneck.

(1) Encoding the basis vectors:

Any basis vector of a $n$-qubit system can be written as $\left\vert
b_{1}b_{2}...b_{n}\right\rangle _{2}=\left\vert b_{1}\right\rangle
_{2}\left\vert b_{2}\right\rangle _{2}...\left\vert b_{n}\right\rangle _{2}$%
\ where $b_{j}\in \{0,1\}$ ($j=1,2,...,n$). To prepare such a vector from
the initial state $\left\vert 0\right\rangle _{2}\left\vert 0\right\rangle
_{2}...\left\vert 0\right\rangle _{2}$ of the quantum circuit, all we need
is to turn the $j$th qubit from $\left\vert 0\right\rangle _{2}$ to $%
\left\vert 1\right\rangle _{2}$\ for any $j$ that satisfies $b_{j}=1$. This
can be done by simply putting a single-qubit NOT gate on the $j$th qubit.
For example, to prepare $\left\vert 00...011\right\rangle _{2}$, we need to
put two NOT gates on the last two qubits. To prepare $\left\vert
11...111\right\rangle _{2}$, we need $n$ NOT gates in total. Therefore, in
average, encoding a single basis vector needs $n/2$ NOT gates.

(2) Encoding the superpositions of two basis vectors:

In this case, we wants to achieve the state $(\left\vert r\right\rangle
_{10}+\left\vert i\right\rangle _{10})/\sqrt{2}=(\left\vert
r_{1}r_{2}...r_{n}\right\rangle _{2}+\left\vert
i_{1}i_{2}...i_{n}\right\rangle _{2})/\sqrt{2}$ where $r_{j},i_{j}\in
\{0,1\} $ ($j=1,2,...,n$). Since $i\neq r$, there exists at least one $j_{0}$
such that $r_{j_{0}}\neq i_{j_{0}}$. Without loss of generality, we can
further assume that $r_{j_{0}}=0$ and $i_{j_{0}}=1$ (if $r_{j_{0}}=1$ and $%
i_{j_{0}}=0$, then the state can be rewritten as $(\left\vert i\right\rangle
_{10}+\left\vert r\right\rangle _{10})/\sqrt{2}$ so that this assumption
still holds). Therefore, we can first prepare the state $\left\vert
r_{1}r_{2}...r_{n}\right\rangle _{2}$ from the initial state $\left\vert
0\right\rangle _{2}\left\vert 0\right\rangle _{2}...\left\vert
0\right\rangle _{2}$, which takes $n/2$ single-qubit NOT gates in average.
Then we put a single-qubit Hadamard gate on the $j_{0}$th qubit, turning its
state into $(\left\vert 0\right\rangle _{2}+\left\vert 1\right\rangle _{2})/%
\sqrt{2}$. Finally, for any $j$ that satisfies $r_{j}\neq i_{j}$ ($j\neq
j_{0}$), we put a two-qubit controlled-NOT gates with the $j_{0}$th
qubit as the control qubit and the $j$th qubit as the target qubit and the
job is done. This generally takes $n/2$ two-qubit controlled-NOT gates in
average, because when taking all possible values of $r_{1}r_{2}...r_{n}$\
and $i_{1}i_{2}...i_{n}$\ into consideration, the probability for finding $%
r_{j}\neq i_{j}$\ is $1/2$. For example, to achieve $(\left\vert
0000\right\rangle _{2}+\left\vert 0110\right\rangle _{2})/\sqrt{2}$, what we
needs is to put a Hadamard gate on the 2nd qubit of the initial state $%
\left\vert 0000\right\rangle _{2}$ so that it becomes $(\left\vert
0000\right\rangle _{2}+\left\vert 0100\right\rangle _{2})/\sqrt{2}$. Then a
controlled-NOT gates with the 2nd qubit as the control qubit and the 3rd
qubit as the target qubit will complete the encoding. Or suppose that we
want to achieve $(\left\vert 0111\right\rangle _{2}+\left\vert
1000\right\rangle _{2})/\sqrt{2}$. The first step is to put three
single-qubit NOT gates on the 2nd, 3rd and 4th qubits of the initial state $%
\left\vert 0000\right\rangle _{2}$, respectively, so that it becomes $%
\left\vert 0111\right\rangle _{2}$. Then a Hadamard gate on the 1st qubit
turns the state into $(\left\vert 0111\right\rangle _{2}+\left\vert
1111\right\rangle _{2})/\sqrt{2}$. Now taking the 1st qubit as the control
qubit, with three two-qubit controlled-NOT gates where the target qubit are
taken as the 2nd, 3rd and 4th qubits, respectively, we eventually arrive at
the state $(\left\vert 0111\right\rangle _{2}+\left\vert 1000\right\rangle
_{2})/\sqrt{2}$.

Actually, in the above process, if we need to put a single-qubit NOT gate on
a specific qubit when turning the initial state $\left\vert 0\right\rangle
_{2}\left\vert 0\right\rangle _{2}...\left\vert 0\right\rangle _{2}$ into $%
\left\vert r_{1}r_{2}...r_{n}\right\rangle _{2}$, and further turning it
into the final state happens to take another two-qubit controlled-NOT gate
with the same qubit as the target qubit (which occurs with probability $1/2$%
), then these two gates can be merged into a single two-qubit negative
controlled-NOT gate (a.k.a. anti-CNOT gate, which keeps the target qubit
unchanged when the state of control qubit is $\left\vert 1\right\rangle _{2}$%
, and applies the NOT operation on the target qubit when the state of
control qubit is $\left\vert 0\right\rangle _{2}$). If this gate is
available as an elementary gate in the quantum hardware, then encoding the
above superpositions of two basis vectors needs only $n/4$ single-qubit NOT
gates and $n/2$ two-qubit gates (including both the controlled-NOT and
negative controlled-NOT gates) in average.

\end{document}